\documentclass[10pt]{article}
\usepackage{indentfirst}
\usepackage[letterpaper]{geometry}
\usepackage{hicss}
\usepackage{times}
\usepackage[none]{hyphenat}
\usepackage{url}
\usepackage{latexsym}
\usepackage{graphicx}
\graphicspath{{images/}}
\usepackage{xurl}
\usepackage{enumitem}

\begin{document}
\setlength\titlebox{5cm}

\title{PyProD: A Machine Learning-Friendly Platform for \\ Protection Analytics in Distribution Systems}

% Comment this for initial manuscript 
% Uncomment this for final manuscript
\author{Dongqi Wu \\
  Texas A\&M University \\
  {\underline{ dqwu@tamu.edu}} \\\And
  Dileep Kalathil \\
  Texas A\&M University \\
  {\underline{ dileep.kalathil@tamu.edu}}\\\And 
  Miroslav Begovic \\
  Texas A\&M University \\
  {\underline{begovic@tamu.edu}} \\\And 
  Le Xie \\
  Texas A\&M University \\
  {\underline{le.xie@tamu.edu}} \\}

\date{}

\maketitle
\begin{abstract}
This paper introduces PyProD, a Python-based machine learning (ML)-compatible test-bed for evaluating the efficacy of protection schemes in electric distribution grids. This testbed is designed to bridge the gap between conventional power distribution grid analysis and growing capability of ML-based decision making algorithms, in particular in the context of protection system design and configuration. PyProD is shown to be capable of facilitating efficient design and evaluation of ML-based decision making
algorithms for protection devices in the future electric distribution grid, in which many distributed energy resources and pro-sumers permeate the system. 

\end{abstract}

\section{Introduction}
%The power distribution grids are undergoing rapid paradigm changes from a predominantly centralized, passive structure toward a grid with many edge-level components such as Distributed Energy Resource (DER), electric vehicle (EV) and battery storage systems. driven by a number of technological, regulatory, and economical factors, there are increasing level of DER penetration at the grid edge.  Policies set forth by federal and state governments such as the the FERC order No.2222 \cite{FERC2222} pave the way for many of the grid-edge level resources to participate and benefit from wholesale market operations. In the past five years, the installed capacity of solar power in Texas have increased exponentially at a rate of more than 50\% annual growth. The recent popularization of EV have put additional stress on the existing power delivery infrastructure as the power capacity of one home EV charger can easily surpass the consumption of an entire house. The addition of  grid-edge components may render conventional planning and operation strategies much less effective. These additional grid elements have brought challenges to the existing distribution system operation methods by creating scenarios that were not previously not considered.  \\

The power distribution grids are undergoing rapid paradigm changes from a predominantly centralized, passive structure toward a grid with many grid edge level components such as Distributed Energy Resource (DER), electric vehicle (EV) and battery storage systems. Such changes are driven by a number of technological, economical and regulatory factors. Recently, the increasingly comfortable driving experience and cheap mileage cost of EVs compared to that of conventional internal combustion engine cars have led to a rapidly growing market. Policies set forth by federal and state governments such as the the FERC order No.2222 \cite{FERC2222} pave the way for many of the grid-edge level resources to participate and benefit from wholesale market operations. In the past five years, the installed capacity of solar power in Texas have increased exponentially at a rate of more than 50\% annual growth. The addition of grid-edge components may render conventional distribution system planning and operation strategies much less effective by creating scenarios that were previously not considered. Specifically in the domain of protection, the presence of DER can invalidate the uni-direction power flow assumption for relay settings and change the pattern of fault current measured at both the fault location and at the substation. The fast charging of EVs can produce current spikes or in general greatly increase the load current which might be mis-recognized as faults. These challenges have to be properly addressed in order to fully realize the potential of grid-edge components.

Recent advances in the computational power of microprocessors and algorithms have led to an increasing popularity of addressing power system challenges with data driven and Machine Learning (ML) methods. However, many of the existing  works provide only a proof-of-concept level solution  to some specific problems on a few tailored small scale test systems. They often lack a thorough and large scale evaluation that include many real-world practical considerations. A major roadblock here is the lack of a public, peer-recognized and professional simulation platform that allow various algorithms, especially machine learning or data-driven methods, to be easily implemented, tested, compared with each other and establish baselines. Most currently available simulators are not designed to accommodate external algorithms, especially ML-based methods, into their simulation process. Moreover, the existing power system simulators demands a significant expertise in power system domain knowledge for testing   algorithms and  interpreting the results. This has become a significant barrier for specialized  data/ML scientists to apply their expertise in solving power system problems. In this paper, we aim to bridge this gap by creating a distribution system simulator that is of comparable accuracy with the state of the art commercial simulators and can be easily integrated with user-developed ML algorithms. We focus on a particular power system application, namely, distribution system protection, as the first step. Our simulation platform PyProD provides a general formulation of ML or data driven approaches to tackle control problems in power distribution studies. It functions as a testbed that can take any system configurations and evaluate abstractly-defined algorithms. In the future, this software will be expanded to go beyond the field of protection and to become a powerful distribution system test-bed for other monitoring and control problems such as voltage/reactive power control, strategic demand response, storage planning and feeder switching.

\subsection{Literature Review}
The various challenges introduced by DERs and Inverter Based Resources (IBRs) for distribution system operators have been extensively studied. Specifically, the current-limiting controls and thermal limit of power transistors in power electronics can greatly reduce their fault current contribution and thus make faults harder to detect for overcurrent relays and fuses \cite{Kou2020} \cite{Walling2008}. Also, the presence of grid-edge distributed generators in general can cause a reduction in fault current measured at the reclosers at the substation which can lead to fault detection and relay coordination failures \cite{Jafari2019}. \\
Many data driven and ML methods has been proposed to boost the performance of fault detection and isolation in distribution systems. Classification algorithms such as Support Vector Machine \cite{Zheng2018}, Bayes classifier \cite{Faiz2007} and Deep Neural Network \cite{Yu2017} \cite{Dalstein1995} are used to improve fault detection and coordination accuracy; Reinforcement Learning (RL) based method \cite{Wu2021} is used to directly control the operation of breaker and reclosers. ML approaches are also popular in other aspects of distribution system operation and control such as reactive power support \cite{ElHelou2021}, storage management \cite{AlSaffar2021} and network switching \cite{Gao2020}. \\
There are a number of existing commercial and license-free power distribution system simulation software. Popular ones include PSS®SINCAL/PSS®CAPE \cite{PSS}, Gridlab-D \cite{GridlabD} and OpenDSS \cite{DSS}. Most software only focus on simulation of the physical distribution grid and provide very limited room for automation and customization. Grid2Op \cite{grid2op} is a pioneering attempt on unifying different classes of algorithms on a common platform and attracting researchers outside the power engineering community to solve power system problems. They built a high-level wrapper around PandaPower \cite{pandapower.2018} and formulated the problem of transmission busbar switching and re-dispatching in an abstract way that is familiar to the ML community. Many successful cross-domain research efforts \cite{Yoon2021}\cite{Lan2019} were made possible by the flexible and open Grid2Op platform. 

\subsection{Contributions}
This paper introduces a open-source simulation platform, PyProD \cite{PyProD}, for the development and evaluation of distribution system protection algorithms. Our software has the following features:
\begin{itemize}
    \item Three-phase unbalanced distribution system fault simulation with dynamic models
    \item Long-term protection reliability assessment using real DER and load profiles
    \item Modular design that allow different custom algorithms to be simulated together
    \item Direct support for ML implementations and automated training with off-the-shelf ML packages
    \item Simple and friendly interface, even for users with little power system domain knowledge
    \item Easily expandable for other studies beside protection 
\end{itemize}

The rest of this paper is organized as follows: Section 2 discusses the design philosophy and architecture of the software, PyProD; Section 3 describes the interfacing with ML/RL tool kits and example implementation of several data-driven algorithms; Section 4 presents several case studies using test networks from different sources with different scales; Section 5 summarizes the paper, propose potential applications and future works.

\section{Structure and Design Consideration of PyProD}

\subsection{Design Guidelines}
There exist several significant obstacles in using  conventional power systems simulation software for developing ML-based solutions for power system problems. First of all, the design objective of most existing power system simulators is mainly on producing trustworthy simulation results on a few representative cases, which is usually sufficient for many conventional power system studies. However,  ML algorithms  require a large amount of simulated data for training their  model as well as for validation. Second, the flexibility of developing, testing and validating user-defined custom algorithms and incorporating them into existing commercial power system simulation packages is extremely hard. %\dc{meaning of expand-ability is not clear. Should we write ``The flexibility of developing, testing, and validating user-defined customized algorithms and incorporating them into the existing simulation platform is extremely hard in  existing commercial power system simulation packages''. You can modify this to convey the message.} 
The source code for these packages were written with a particular priority on improved computation speed since the need for customized algorithms  was not imminent at the time of their development. Researchers who wish to develop and test new algorithms on simulators often need to write complicated wrapper functions around the limited input/output interface of pre-compiled binaries, or carefully dive into source codes to insert custom code blocks into simulation streamline. As a result, it is very difficult to replicate and evaluate different proposed algorithms on a common test-bed, as each of them require dedicated implementations that are not compatible with each other. Third, almost all  ML methods are built upon various libraries written in scripting languages, mostly Python. The lack of native scripting support of most of the existing power system simulators has made interfacing with advanced ML packages   difficult and inefficient. Fourth, deep domain knowledge on power systems is usually required for users to understand and interpret simulation results in existing power systems simulation tools. This pauses a major hurdle of ML researchers without extensive training in power system analysis in contributing to solving power systems problems through ML methods.   
% often find it difficult to actively contribute with the help of power system specialists.

The software PyProD is designed with the goal of  addressing the aforementioned problems. We aim to develop a simulation platform that is capable of producing trustworthy simulation results while at the same time cater to the needs of various algorithm developers to the maximum extent. Specifically, we propose to incorporate following aspects to improve compatibility and accessibility of power distribution system simulator for protection design:
\begin{itemize}[leftmargin=\parindent]
    \item \textbf{Creation of large training dataset:} We create a large amount of fault scenarios by combining realistic wind/solar/load profile with Monte-Carlo fault simulation. Each scenario contains a time window of simulation around a fault with random parameters including location, time, type and impedance. The initial condition of the simulation is based on a sample draw from the profile data-set which defines the load capacity and output level of DER generators.
    \item \textbf{Interfacing with user-developed algorithms:} The software is developed in a modular fashion such that the physical grid simulator is configured to automatically exchange data with independently defined custom controllers throughout the simulation process. The simulation can be performed  with or without an arbitrary number of user-written controllers. As a result, the users are able to focus on the algorithmic aspect alone, without worrying about the power system interface aspects.
    \item \textbf{Programming language compatibility:} To be adaptive to algorithms based on different software packages, especially with existing ML packages, the entire software is written in Python and the simulation process can be fully controlled by Python scripts. 
    \item \textbf{Research community accessibility:} The simulation platform is configured based on the widely accepted OpenAI Gym  \cite{gym} and the relevant variables in simulation results are interpreted, visualized and put into context that is familiar to the ML community. This allows researchers with little power system background to understand the problem and bootstrap using one of the many off-the-shelf RL packages that are built for OpenAI Gym environments.
\end{itemize}

\begin{figure*}[ht]
\centering
\includegraphics[scale=0.45]{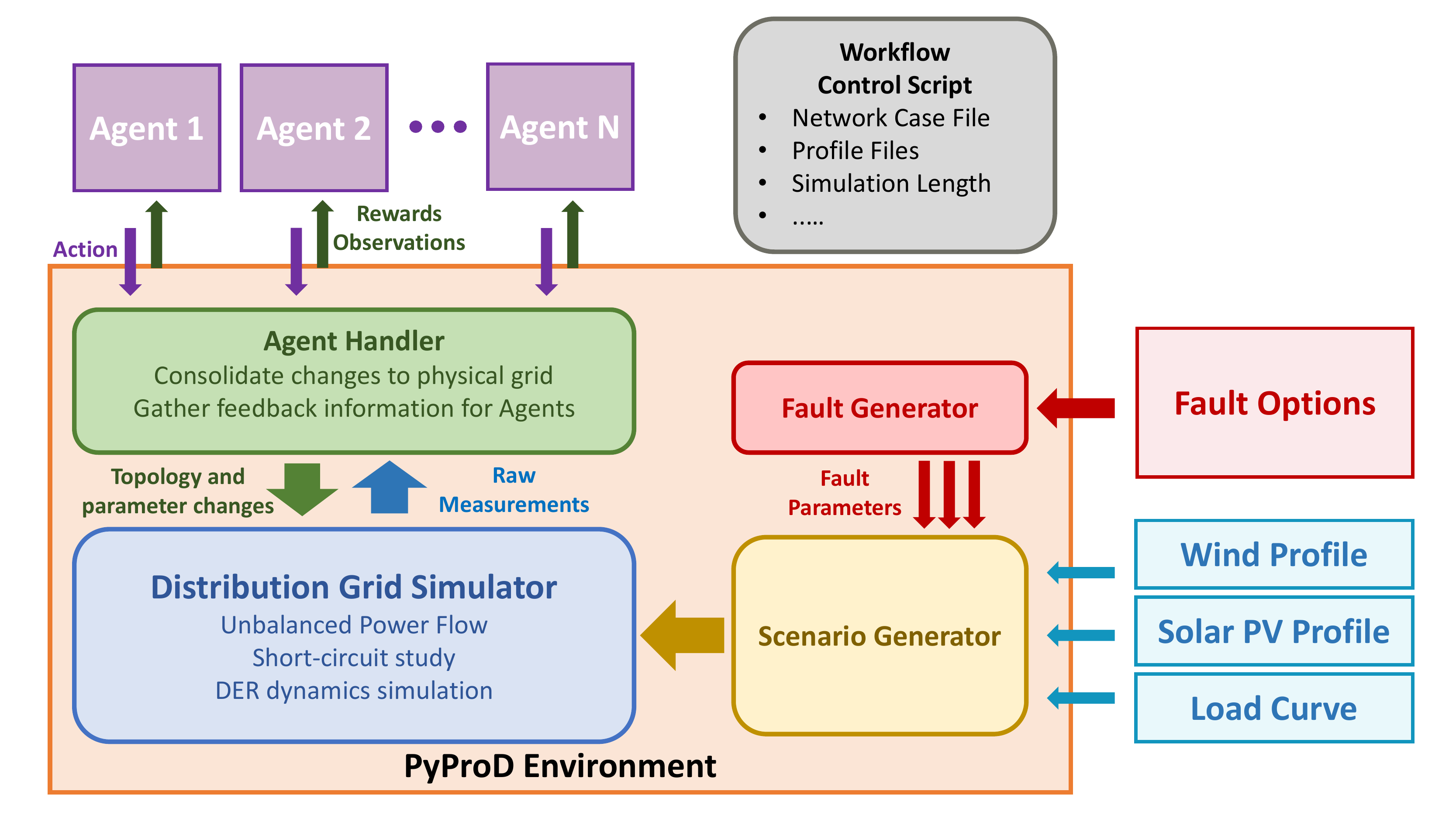}
\caption{Architecture of PyProD}
\label{fig:flowchart}
\end{figure*}

\subsection{Software Architecture and Implementation}

A conceptual flowchart of PyProD is shown in Fig. \ref{fig:flowchart}. The PyProD environment is the main engine where simulations and case studies are performed. The environment itself can be run as a standalone simulator to perform simple short-circuit studies and visualize results. The simulation is organized into independent \textit{episodes} with a limited number of time-steps. Each episode represents a fault scenario in which the simulation starts from a steady state, then later a fault or disturbance is added and the simulation continues for a certain time to observe the system dynamic and protection operations. The user can optionally supply profile data that defines the initial power flow of each episode. The \textit{Scenario Generator} can be configured to do month/year-long consecutive simulation or randomly draw samples from the profile dataset. Similarly, the user can control the desired type and range of parameters, within which random faults are created, in the \textit{Fault Generator}. Note that each module described above can run independently if the user chooses to skip some blocks and perform simulation manually instead. 

Each protective device (relay, recloser, fuse) is represented as an \textit{Agent}, which interact with the environment by observing grid states (e.g. voltage, current, apparent impedance, etc.) and attempt to respond to fault by switching off/tripping grid components. To account for multiple agents running in parallel, the \textit{Agent Handler} goes through all active agents, provide the observation they need and collect their action outputs at the beginning of each simulation step. The type of measurements and possible control outputs (a.k.a state and action space) of agents are defined within the \textit{Agent} framework. Throughout the simulation, the \textit{Agent Handler} will automatically push the required measurements and update the state variables of every \textit{Agent}. The users only need to implement their custom algorithm abstractly by creating a function that maps observations to actions.

The main PyProD environment is designed around the OpenAI Gym \cite{gym} and can be used exactly like one of the standard RL training and testing environments. This framework has important advantages that are not available for conventional power systems  simulators. Most importantly, it allows the many publicly available RL packages to be used directly in the development of new algorithms for power systems problems.  Packages such as stable-baseline \cite{stable-baselines3} and TensorForce \cite{tensorforce}  are commonly used by the RL community as they contains some of the best performing and state-of-the art  RL algorithms and the supporting build-blocks. These packages are all designed to work within the Gym standard. Researchers in the control and learning community are familiar with the Gym framework and they usually develop and testing their algorithms using this framework. Adopting the same standards and framework make the PyProD platform more friendly and appealing for ML researchers.

\noindent  \textbf{Distribution Grid Simulator}\\
The core of PyProD is the simulator that perform power flow and short circuit analysis. The 3-phase unbalanced power flow and fault simulation in PyProD are performed using OpenDSS\cite{DSS}, which is a well recognized and reliable grid simulator that have been used in many research works. We choose OpenDSS to be the core engine as it has several unique advantages over other comparable simulators. First, OpenDSS can be controlled by sending commands through a COM server in programming languages, which is a necessary feature for automation. Second, OpenDSS supports dynamic models for Photovoltaic (PV) and induction generators. In distribution systems where generators are mostly from renewable energy sources, it is essential to use the appropriate models. Third, there are a variety of existing public test feeder case files of different size in DSS format that allows quick testing over a large selection of networks. Finally, PyProD provides a set of easy-to-use Application Programming Interfaces (APIs) for interacting with OpenDSS as the default way of data exchange through the COM interface is difficult to use.

\noindent \textbf{Data Source and Scenario Creation}\\
To satisfy the data size requirement for data-driven and learning-based methods, it is essential to efficiently produce a large amount of meaningful fault scenarios. In PyProD, each scenario consists of an unique combination of power flow pattern and a fault in the network. OpenDSS dynamic simulation is used to produce the transient response after the fault. The network steady-state powerflow to initialize each episode is determined by the load distribution and output of distributed generators, if they exist in the network. All profiles are provided in the format of percentage, which will be used to scale the base values  in the original DSS case file. PyProD comes with a selection of profile dataset from publicly available hourly load and wind/PV generation data disclosed by several major Regional Transmission Organizations (RTOs) in the United States. During each episode, a fault is added to the network to test the operation of protection devices. This fault is set to have randomized parameters including type, location, fault impedance and occurrence time. Since the core engine, OpenDSS is an unbalanced 3-phase simulator, its possible to create scenarios involving faults with different connection between the phases, neutral and ground nodes at each bus. PyProD automatically scan through the network to determine valid location of faults. Since OpenDSS only allow faults to happen at buses, a dummy bus needs to be added temporally for faults in the middle of lines or other grid components. For distribution systems without neutral grounding across the entire network (e.g. IEEE 37 bus feeder), PyProD will only simulate two or three phase faults. Other disturbances such as generator tripping and line opening can also be included if desired. The user can set the range of each parameter within which faults are created.

\section{Interfacing Distribution System Simulation with Machine Learning}

\subsection{Environment and Agent Interaction}
\begin{figure*}[ht]
\centering
\includegraphics[scale=0.45]{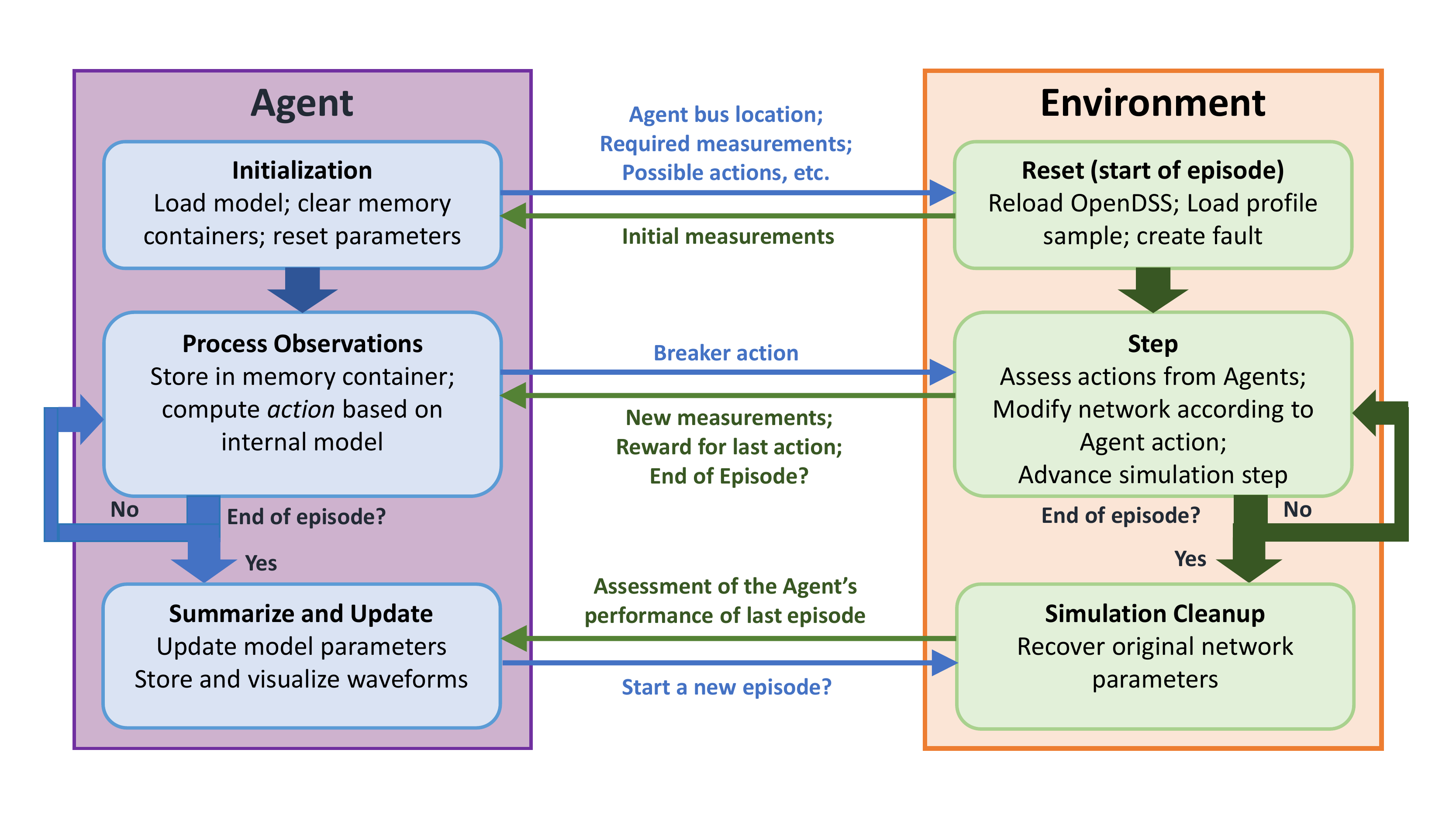}
\caption{Interaction Between Agent and Environment}
\label{fig:agent}
\end{figure*}

Similar to other standard RL environments, the controllers, or protection devices under this context, are modelled as \textit{Agents}. Each Agent retrieve observations from the environment and output actions that are determined by its model. A substantial difference between PyProD and other Gym-like environments is that PyProD is designed to be a multi-agent platform as there are often multiple protection devices working together in a single distribution network. Under the PyProD framework, each agent is a self-contained script segment that communicate with the environment through a few designated functions. Figure \ref{fig:agent} shows a diagram of data flow between an Agent and the environment.

PyProD provides an \textit{Abstract Class} definition for agents as template. The simplest form of an agent needs to have the following two functions: \textit{observe} and \textit{act}. They will be automatically called by the environment during the simulation to facilitate data exchange between the environment and the agent. Beside the basic functions, an agent may have as much internal state variables or function as needed, they are not visible to the environment. The \textit{observe} function takes an object of a helper class \textit{dssCase}, which include the standard handle variable to access OpenDSS through the COM object and default APIs, plus many functions that allow easy retrieval of simulation information. Within the \textit{observe} function, The agent needs to gather necessary measurement and store to an internal container. For example, the \textit{observe} function for a basic overcurrent relay would simply be collecting current magnitude at the branch it operates from and store it as a variable. The \textit{act} function computes the breaker action (whether to trip or not) and output a single binary to the environment. Within the agent, the action can be computed based on either a simple immediate observation or a complicated multi-step state transition model, the environment only take the breaker action. For a instantaneous overcurrent relay, the \textit{act} function will simply be a condition checking if the fault current is larger than the pickup threshold. A set of ready-to-use agents modelling various conventional protection relays such as inverse-time overcurrent, differential and distance relays are available with the installation along with example execution scripts. This agent and environment interaction framework can also be easily extended to model the control of other types of grid components. For example, a voltage regulator can be modelled as an agent that takes the voltage measurements as observation and outputs a winding tap position. 

\subsection{Example: Sequential RL and SVM Based Relay}
ML based agents differ from conventional agents mainly in the aspect that they need to be trained using a large amount of data generated from simulations, while conventional ones take a set of analytically determined parameters. PyProD can be used to automatically produce a large training dataset by running Monte-Carlo simulation. In the following the procedures and components used to create ML based agents are illustrated using two types examples: a Reinforcement Learning (RL) based relay and supervised learning based relays using Support-Vector-Machine (SVM) and Deep Neural Network (DNN).

\begin{figure}[ht]
\centering
\includegraphics[scale=0.23]{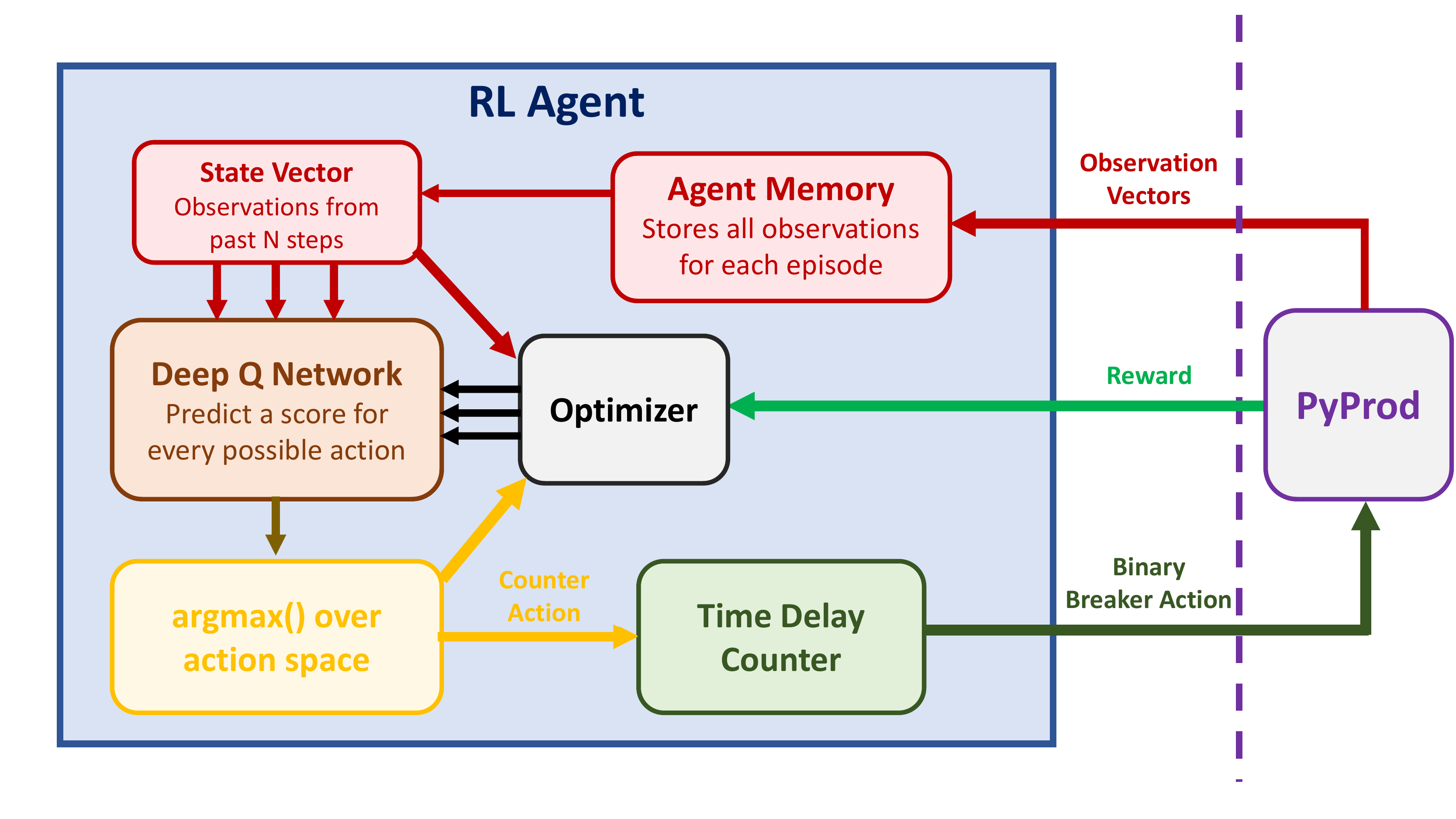}
\caption{Conceptual Diagram of an RL Relay Agent}
\label{fig:rlrelay}
\end{figure}

\noindent \textbf{RL Relay Implementation}\\
RL is a major subsidiary of machine learning that focus on controlling dynamic systems. The environment to be controlled is modelled as a Markov Decision Process with an unknown state transition kernel. In \cite{Wu2021} an RL formulation of protective relay control using Deep-Q-Network (DQN) is discussed in detail. Generally, training an RL relay requires an responsive environment with which the in-training agent can try different actions and receive resulting observation and evaluation for the actions the agent choose. This evaluation is done through passing an \textit{Reward} signal from the environment to the agent along with new observations for each action. The DQN learns to select actions that maximize the expectation of total reward obtained in every episode. Hence, the environment should return a high reward for correct actions (e.g. tripping after fault in zone) and a low reward for incorrect actions (e.g. no operation after fault, tripping the line when there is no fault). PyProD automatically analyze the operation of relays to see if the actions are desirable and return the corresponding reward values. In a multi-agent formulation, PyProD performs graph searching for each fault scenario to determine the coordination relationship of all relays in the systems to make sure that coordinated backup operations are rewarded appropriately. There are several different reward designs the comes with the PyProD installation as example. A diagram showing the structure of an RL agent is shown in Fig. \ref{fig:rlrelay}.

\begin{figure}[t]
\centering
\includegraphics[scale=0.23]{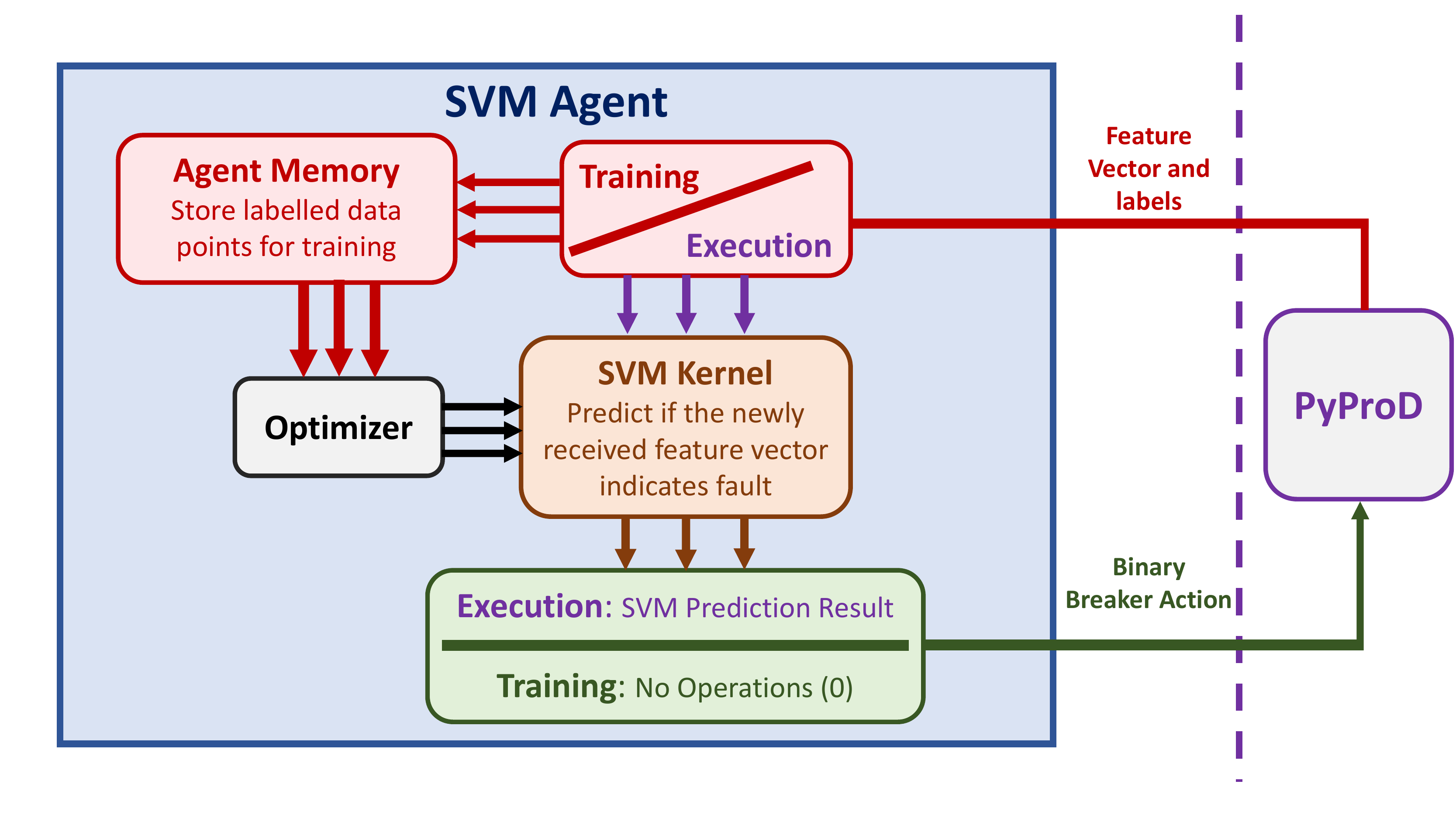}
\caption{Conceptual Diagram of a SVM Relay Agent}
\label{fig:svmrelay}
\end{figure}

\noindent \textbf{SVM Relay Implementation}\\
Support-Vector-Machine (SVM) \cite{svm} is a supervised learning model for the classification of labelled data points. SVM and its variations are considered to be among the most robust and widely used classification algorithms. In SVM, each input data point is considered a point in a high dimensional space in which the dimension equals the number of features of each data point. The features may include local voltage/current (in peak, sequence or phase values) and frequency measurement. The classifier is computed by calculating the \textit{maximum-margin hyperplane} between two groups of labelled training data. SVM and other supervised learning methods require a large number of data points in the training dataset. This can be directly obtained by using the Monte-Carlo simulation function in PyProD. The function runs randomly generated fault scenarios until the user-specified size of training data is collected, the data points are automatically labelled based on the desired operation of the relay to be trained. In the validation stage, new data points are created and run through the SVM model to test for accuracy. The distribution of training and validation dataset creating can be controlled though changing the load and generation profile data. A diagram of a SVM relay is shown in Fig. \ref{fig:svmrelay} and the implementation is provided with PyProD.

\noindent \textbf{DNN Relay Implementation}\\
Deep Neural Network (DNN) is another well-known classification method based on the supervised learning concept. Similar to that of SVM, the training of DNN model takes a collection of labelled high-dimensional data points and optimizes over the weights and biases of the neural network model to produce the best possible classification accuracy in the training dataset. In theory, A DNN model is more flexible than SVM as deep neural networks can approximate any function \cite{Hornik1989} even for those with high non-linearity. More sophisticated variations such as Convolutional Neural Networks and Recurrent Neural Networks can be used to extract spatial and temporal, respectively, coherence of input features. We have implemented a simple DNN based agent that uses the positive sequence voltage and current of the past 10 simulation steps to determine whether a fault is present in its downstream area. The structure, training process and implementation of a DNN agent is similar to the SVM agent except for that the kernel is replaced by a deep neural network.

\section{End-to-End Case Studies}
In this section, we demonstrate the end-to-end usage of PyProD by performing case studies on three distribution networks of different origin, configuration and dimension. PyProD is able to appropriately handle a wide range networks in performing simulation, training learning-based agents for each network and evaluating their performance. 

On each test case, we compare the performance of the following four types of relay control algorithms: 1) Traditional inverse-time overcurrent (OC) relay; 2) Supervised learning using SVM; 3) Supervised learning using deep neural network; 4) Reinforcement learning using deep-Q-network. The pickup current threshold of the inverse-time overcurrent agent is set to be two times the maximum possible load current (with largest load capacity and minimum distributed generation). The parameters of the ML based agents are trained automatically using the same agent scripts to emphasize the flexibility of the PyProD framework.

The performance of agents are evaluated by running many random episodes. Each episode may contain a fault that one or more agent need to respond to. Depending on the operation of the agents, the episodes can have one of the following outcomes: 1) correct, if all agents performed the desired operation throughout the episode; 2) false positive, if one or more agent tripped when there is no fault; 3) false negative, if one or more agent fails to trip when a fault is in its region; 4) coordination failure, if an agent tripped when the fault is outside its region, or the backup agent tripped after a fault before the closest agent did. The total success rate is determined to be the percentage of correct episodes.

\subsection{IEEE 34 Bus System}
\begin{figure}[t]
\centering
\includegraphics[scale=0.6]{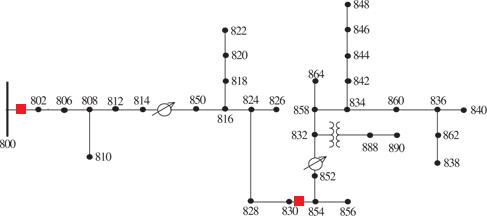}
\caption{The IEEE 34 Node Feeder}
\label{fig:ieee34}
\end{figure}

The IEEE 34-bus feeder \cite{IEEEfeeder} in Fig. \ref{fig:ieee34} has been recognized as one of the most common and popular benchmark system in distribution system studies. It is a long feeder that has a combination of three phase lines and several single-phase laterals. We use this network to test the impact of DER on fault detection and relay coordination. Specifically, we add a few DERs to the system as listed in Table \ref{table:config34}. Two protective relays are placed at the substation (bus 800) and bus 830, which is right after the long 6.23 km line 828-830 and is the beginning of the second half of the circuit. When faults occur to the downstream of bus 830, the relay at 830 is expected to operate first and the relay at bus 800 should operate later with a sufficiently long delay. The load and DER profile for simulation on the IEEE 34 bus feeder is adopted from the demand data from 
the South hub of Southwest Power Pool (SPP) and the SPP Fuel Mix report respectively.

\begin{table}[h]
\begin{tabular}{|c|c|c|c|c|}
\hline
Bus & Phase & Model & Rated kVA & Connection \\ \hline
846 & ABC   & PV    & 200       & wye        \\ \hline
820 & A     & PV    & 50        & /          \\ \hline
838 & B     & PV    & 50        & /          \\ \hline
860 & ABC   & Wind  & 200       & delta      \\ \hline
814 & ABC   & Wind  & 200       & delta      \\ \hline
\end{tabular}
\caption{DER Placements in the IEEE 34-Bus Feeder Case}
\label{table:config34}
\end{table}

Each agent uses an appropriate method to facilitate the coordination between two relays: The RL agent can learn the optimal delay for each scenario as a part of its formulation; The operation time of inverse-time overcurrent relays is determined by the IEEE Very Inverse curve; The two supervised learning based, SVM and deep neural network respectively agents utilizes multi-class classification to choose from the following three options: 1) no trip; 2) trip instantly; 3) trip 0.3 seconds later. In SVM this is realized through an one-vs-all approach which trains a classifier for each option. In the deep neural network approach, the model is trained to give correct actions the highest score in a 3 neuron output layer using the \textit{softmax} activation. The parameters of the overcurrent agents are calculated from the extreme cases obtained in running a yearly simulation using load and DER profile data. The size of training dataset for the 3 learning based agents (RL, SVM and Deep neural network) are all limited to 500 random episodes. The performance of the agents are quantified by the number of correct episodes in a validation run of 2000 random episodes. The performance of the four type of agents is summarized in Table. \ref{table:der34}.

\begin{table}[h]
\centering
\begin{tabular}{|c|c|c|c|c|}
\hline
Agent       & \begin{tabular}[c]{@{}c@{}}False\\ Positive\end{tabular} & \begin{tabular}[c]{@{}c@{}}False\\ Negative\end{tabular} & \begin{tabular}[c]{@{}c@{}}Coord.\\ Failure\end{tabular} & \begin{tabular}[c]{@{}c@{}}Success\\ Rate\end{tabular} \\ \hline
OC       & 0  & 321 & 66 & 80.65\%  \\ \hline
RL       & 0  & 28  & 6 & 98.3\%  \\ \hline
SVM      & 2  & 19  & 75 & 95.2\% \\ \hline
Deep-NN  & 26 & 12  & 47 & 95.75\%  \\ \hline
\end{tabular}
\caption{Performance of Different Agents on IEEE34 with DER in 2000 Episodes}
\label{table:der34}
\end{table}

It can be seen that the simple overcurrent design have suffered from the addition of DER at the load side of the feeder. In many cases the fault current have dropped below the pickup current or the fault current contribution from DERs have caused coordination problems. The two supervised learning methods (Deep-NN and SVM) have similar performance in fault detection, but their coordination performance in inferior to the RL formulation which, in additional to detection, is also designed to learn the optimal tripping delay for each fault scenario.

\subsection{IEEE 37 Bus System}
\begin{figure}[t]
\centering
\includegraphics[scale=0.7]{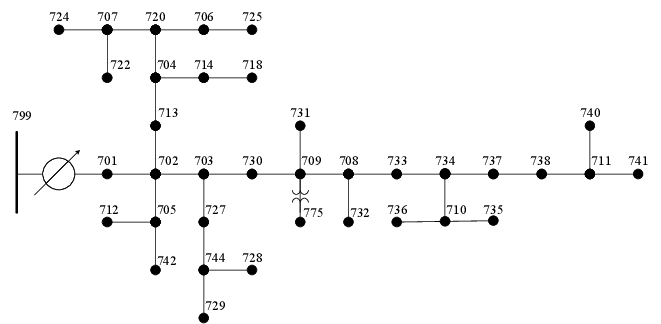}
\caption{The IEEE 37 Node Feeder}
\label{fig:ieee37}
\end{figure}

The IEEE 37-bus feeder \cite{IEEEfeeder} shown in Fig. \ref{fig:ieee37} represents a real distribution system in California that operates at 4.8 kV. The feature of this system is that there is no neutral grounding across the circuit and all loads are connected in delta. In such configurations single phase to ground (SLG) faults cannot produce a significant short-circuit current as there is no closed circuit for the fault current though ground. Protective relays are not expected to operate since the system is expected to be operational for a extended period. PyProD can automatically identify the circuits that do not have a ground path and adjust the simulation accordingly to avoid mistakenly labelling agents not responding to SLG faults as failed operations. 
In this case we consider a simpler single-agent case where there is no DERs and one relay is placed at the substation (bus 701). Hence, the variation among scenarios only involves the variations in load profiles and fault parameters. The load profile is adopted from the 2019 Los Angles demand record retrieved from CAISO. A similar experiment process is performed in PyProD and the results are shown in Table \ref{table:ieee37}.

\begin{table}[h]
\centering
\begin{tabular}{|c|c|c|c|}
\hline
Agent       & \begin{tabular}[c]{@{}c@{}}False\\ Positive\end{tabular} & \begin{tabular}[c]{@{}c@{}}False\\ Negative\end{tabular}  & \begin{tabular}[c]{@{}c@{}}Success\\ Rate\end{tabular} \\ \hline
OC       & 0  & 359  & 82.05\%  \\ \hline
RL       & 0  &  11 & 99.45\%  \\ \hline
SVM      & 0 &  51 & 97.45\% \\ \hline
Deep-NN  & 4 & 23  & 98.65\%  \\ \hline
\end{tabular}
\caption{Performance of Different Agents on IEEE37 in 2000 Episodes}
\label{table:ieee37}
\end{table}

In this single-agent case, the three learning-based method have shown to be quite accurate in the detection of faults. The detection performance of the overcurrent agent is affected by the fault impedance, as all the false-negatives are faults whose fault impedance ranges from 6 to 15 ohms.

\subsection{Austin Distribution Systems}
The combined transmission \& Distribution synthetic grid dataset Syn-Austin-TDgrid-v03 \cite{Li2020} has a collection of realistic, high resolution and large-scale distribution feeder network models. The dataset has a large collection of 448 feeders distributed among 140 substations in the area surrounding Austin, Texas. For these large and complex feeders, PyProD is able to easily integrate and test abstractly defined ML algorithms in simulations. Similarly, the performance of the 4 different agents is listed in Table \ref{table:austin}.

\begin{table}[h]
\centering
\begin{tabular}{|c|c|c|c|}
\hline
Agent       & \begin{tabular}[c]{@{}c@{}}False\\ Positive\end{tabular} & \begin{tabular}[c]{@{}c@{}}False\\ Negative\end{tabular}  & \begin{tabular}[c]{@{}c@{}}Success\\ Rate\end{tabular} \\ \hline
OC       & 0  & 148  & 92.6\%  \\ \hline
RL       & 0  &  66 & 96.7\%  \\ \hline
SVM      & 0 &  124 & 93.8\% \\ \hline
Deep-NN  & 41 & 76  & 94.15\%  \\ \hline
\end{tabular}
\caption{Performance of Different Agents on a Synthetic Austin Feeder}
\label{table:austin}
\end{table}

\begin{figure}[t]
\centering
\includegraphics[width=0.5\textwidth]{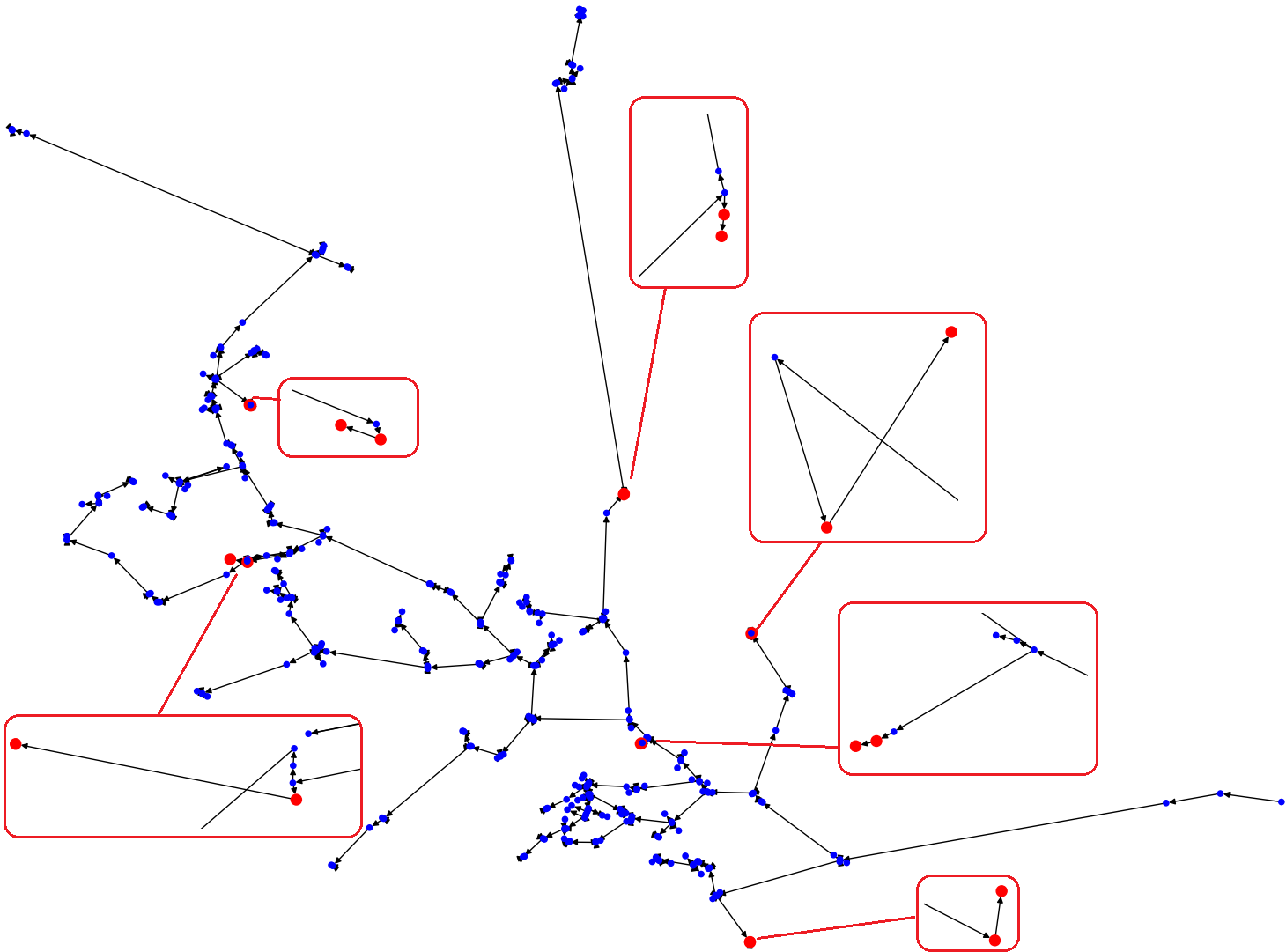}
\caption{Diagram of a Synthetic Austin Feeder with Buses Under-reaching}
\label{fig:austin}
\end{figure}

We use one of the Austin synthetic distribution feeder to demonstrate the visualization capability of PyProD for distribution system analysis, as well as its ability to handle large and complicated network. This particular feeder has 1417 buses and 1244 lines including fuses and switches. When creating the environment for a OpenDSS case file, PyProD attempts to parse the circuit as a directed graph with NetworkX \cite{networkx}. In addition to the graph traversal algorithms, we also expand on the visualization tools to present simulation results. A common problem for the protection of large and long feeders is that the recloser at the substation is usually less sensitive to faults deep into the circuit (a.k.a. under-reaching). This is because a large equivalent impedance between the substation transformer and the grounding point can reduce the magnitude of fault current, which can cause the recloser failing to detect the fault, or a large reclosing delay during which the fuses within the feeder might melt unnecessarily. PyProD is able to identify the buses that are susceptible to causing under-reaching problem to an agent during simulation. Fig. \ref{fig:austin} shows the diagram of one feeder from the synthetic Austin dataset. We conducted a Monte-Carlo fault simulation using year-round load profile from Houston, TX and simulated faults with 0 ohm impedance. The blue dots marks the buses on which all faults are successfully detected by an overcurrent agent at the substation, where the red dots marks the buses on which at least one simulated fault is not detected. As shown in the graph all the under-reaching faults are concentrated to only a few laterals. Hence, it might be helpful to put additional fuses in these laterals to improve reliability.

%\subsection{EPRI Ckt5}
%\redtext{do we need another case study on the EPRI Ckt5 case?}

\section{Concluding Remarks}
We develop an ML friendly simulation platform, PyProD, for electric distribution system protection  simulation and analysis. PyProD is designed to bridge the gap between conventional electric distribution grid simulation and various machine learning algorithms. It serves researchers in both communities as an open platform to analyze, compare, and propose new design algorithms for protection systems of the future.  PyProD is built around the OpenAI Gym architecture which is standard in the ML/RL  community, and can be directly used with many machine learning packages. It decouples the physical grid simulation and control algorithm design by constructing the protective relays as external agents that interact with the grid. This allows developers, even with little power system domain knowledge, to design and evaluate control algorithms abstractly without the need to operate the power system simulation software. PyProD is capable of using historical load and renewable generation profiles to generate a large amount of simulated data for the training and testing of ML and conventional protection strategies.

In the future, we plan to expand PyProD to include other types of distribution system monitoring and control problems that has the potential to be transformed by  methods. We will also go beyond the capability of OpenDSS to develop more realistic models of grid-edge components such as solar/wind generators, battery storage and electric vehicles. Another direction will be designing and implementing a market mechanism to study possible economic benefits of flexible loads and strategic demand response from the perspective of individual or aggregated customers.

\section{Acknowledgement}
This work is supported in part by Texas A\&M Smart Grid Center, and in part by NSF OAC-1934675, CMMI-2130945.

%if added before the last page, this command can help balancing columns
%\addtolength{\textheight}{-.2cm} 

%Bibliography 
\bibliographystyle{ieeetr}
\bibliography{sample}

\end{document}